\documentclass[%
 reprint,
superscriptaddress,
 amsmath,amssymb,
 aps,
floatfix,
]{revtex4-2}

\usepackage{graphicx}
\usepackage{dcolumn}
\usepackage{bm}
\usepackage{hyperref}

\usepackage{color}

\usepackage{soul}

\begin{document}


\title{Broadly tunable photon pair generation in a suspended-core fiber}

\author{Jonas Hammer}
\email{jonas.hammer@mpl.mpg.de}
 \affiliation{Max-Planck-Institute for the Science of Light, Staudtstra\ss{}e 2, 91058 Erlangen, Germany}%
 \affiliation{Physics Department, Friedrich-Alexander University, Staudtstra\ss{}e 2, 91058 Erlangen, Germany}

\author{Maria V. Chekhova}%

 \affiliation{Max-Planck-Institute for the Science of Light, Staudtstra\ss{}e 2, 91058 Erlangen, Germany}%
 \affiliation{Physics Department, Friedrich-Alexander University, Staudtstra\ss{}e 2, 91058 Erlangen, Germany}
\author{Daniel R. H\"aupl}

 \affiliation{Max-Planck-Institute for the Science of Light, Staudtstra\ss{}e 2, 91058 Erlangen, Germany}%
 \affiliation{Physics Department, Friedrich-Alexander University, Staudtstra\ss{}e 2, 91058 Erlangen, Germany}
\author{Riccardo Pennetta}
 \affiliation{Max-Planck-Institute for the Science of Light, Staudtstra\ss{}e 2, 91058 Erlangen, Germany}%
\author{Nicolas Y. Joly}
 \affiliation{Physics Department, Friedrich-Alexander University, Staudtstra\ss{}e 2, 91058 Erlangen, Germany}
 \affiliation{Max-Planck-Institute for the Science of Light, Staudtstra\ss{}e 2, 91058 Erlangen, Germany}%

\date{05.11.2019}


\begin{abstract}
Nowadays fiber biphoton sources are nearly as popular as crystal-based ones. They offer a single spatial mode and easy integrability into optical networks. However, fiber sources lack the broad tunability of crystals, which do not require a tunable pump. Here, we report a broadly tunable biphoton source based on a suspended core fiber. This is achieved by introducing pressurized gas into the fibers hollow channels, changing the step index. The mechanism circumvents the need for a tunable pump laser, making this a broadly tunable fiber biphoton source with a convenient tuning mechanism, comparable to crystals. We report a continuous shift of 0.30\,THz/bar of the sidebands, using up to 25\,bar of argon.
\end{abstract}

\maketitle

Optical fibers are an ideal platform for the generation of entangled photon pairs (biphotons) via spontaneous four-wave-mixing (FWM), due to their long light-matter interaction length. In particular, solid core fibers offer high effective nonlinearity. However, they typically lack the convenient tuning mechanism present in crystal-based biphoton sources, beside the trivial but costly scheme of tuning the pump wavelength. Biphoton and twin-beam emission from crystal sources can be easily tuned over bandwidths exceeding 100\,THz via angle tuning. This property has recently been exploited as a means for absolute spectrometer calibration \cite{Lemieux2019}. Biphotons tunable over $\sim 15$\,THz from a crystalline (lithium niobate) whispering gallery mode resonator have been demonstrated by heating the resonator \cite{Foertsch13}. Silicon waveguides have gained popularity as sources of entangled photon states \cite{Feng19}, and tunability might be implemented using temperature tuning \cite{Wang16}.
In fibers, a  limited amount of tuning (few THz) has been demonstrated by stretching or heating \cite{Ortiz-Ricardo17}, however, these approaches are limited by fiber damage.
Meanwhile, gas filled hollow-core fibers offer broad dispersion tuning, but implementing a biphoton source in these fibers remains a challenging task, due to their low nonlinearity.
Here, we combine the high nonlinearity of a solid core fiber with the convenient tuning scheme offered by gas filled fibers, to implement a tunable source of entangled photons. This is achieved by filling the channels surrounding the core of a suspended-core fiber (SCF).
SCFs are a class of index-guiding microstructured fibers, where light is guided in a glass core suspended by three glass nano-membranes. SCFs have been used in a variety of applications, ranging from supercontinuum generation~\cite{Fu08} to gas absorption spectroscopy~\cite{Webb07}, or chemical sensing~\cite{Cubillas13}. 

In FWM, two photons of an incident beam (denoted pump) are annihilated and the energy is transferred to two daughter photons (denoted signal and idler) symmetrically spaced around the pump. 
The signal and idler frequencies $\omega_\text{S}$,  $\omega_\text{I}$ are determined by the phase-matching conditions:
\begin{align} \label{eq:PM}
       \Delta\beta = 2\beta(\omega_\text{P}) - \beta(\omega_\text{S}) - \beta(\omega_\text{I}) -2 \gamma P_\text{Pump} &=0 \,\, \\ 
       2\omega _\text{P} - \omega_\text{I} - \omega_\text{S} &= 0 \, ,\label{eq:Energy}
\end{align}
where $\omega_\text{P}$ is the pump frequency and $\beta$ denotes the propagation constant. $P_\text{Pump}$ is the peak power of the pump beam, and $\gamma$ is the effective nonlinearity of the fiber mode. The last term in Eq.~\ref{eq:PM} corresponds to a nonlinear contribution to the phase-mismatch due to cross-phase and self-phase modulation terms. Here, we refer to the blue shifted sideband as signal, and the red shifted one as idler.  Eq.~\ref{eq:Energy} ensures energy conservation and leads to a symmetric frequency detuning of the two sidebands.

 If a seeding beam at either of the sideband frequencies is present, FWM is an entirely classical phenomenon, also known as parametric amplification. It is exploited in fiber-optic parametric amplifiers or oscillators~\cite{Pocholle85,Hansryd02}. If no seeding beams are launched, FWM should be described in the framework of quantum mechanics. In this case the process can be thought of as seeded by vacuum fluctuations and produces photon pairs at the sideband frequencies. This has been used as a source of biphotons in dispersion shifted~\cite{Li04,Takesue04}, microstructured~\cite{Rarity05,Fan05} or polarization maintaining fibers~\cite{Smith09}. Recently, optical fiber tapers have gained some popularity as biphoton sources~\cite{Kim19,shukhin19}. Fiber tapers exhibit dispersion and guidance properties similar to SCF, however, the additional structure surrounding the light-guidance region makes SCF less prone to damage and environmental influences.
 A tunable source of bright squeezed-vacuum twin beams based on a FWM process (modulational instabillity) has been demonstrated in hollow-core fibers~\cite{Finger15}. For such a source, tunability of the number of modes has been demonstrated~\cite{Finger17} and proposed for further use in a tunable biphoton source~\cite{Cordier19}. 

Here, we use a similar tuning principle but apply it to SCF, where the solid core provides a high nonlinearity. First, we demonstrate the continuous tuning of seeded FWM. Using parametric amplification of a broadband infrared seed, we observe a shift of $7$\,THz for the emitted conjugate near-infrared sideband ($\sim 17$\,nm at this wavelength range), when filling the fiber with up to 25\,bar of argon. We then demonstrate a fiber biphoton source with very high coincidence-to-accidental ratio (CAR) and tunability over at least $4.6$\,THz ($\sim 30$\,nm measured at the red detuned wavelength, we estimate this shift can be  $\sim40$\,THz if 45\,bar of xenon are used as filling gas). We measured a coincidence rate of $76\,$s$^{-1}$. Accounting for losses due to spectral filtering and fiber coupling, as well as detection efficiencies, this corresponds to a pair production rate on the order $10^3\,$s$^{-1}$ at 40\,mW CW pump power.

\begin{figure}[htbp]
    \centering
    \includegraphics[width=\linewidth]{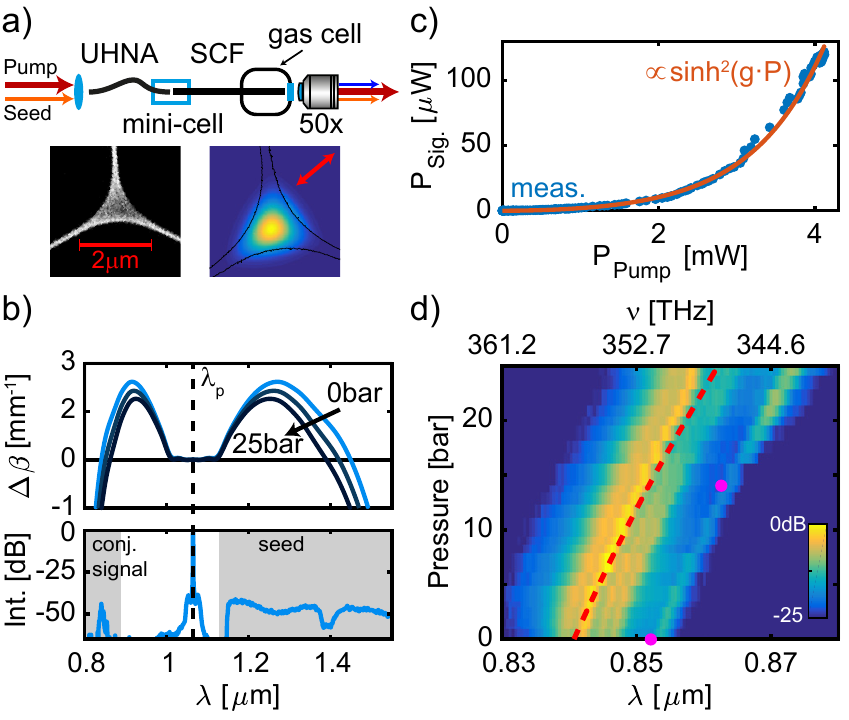}
    \caption{Seeded FWM. a) Setup for pressure tuning the FWM, see text. The inset shows a scanning electron micrograph of the SCF core-region (left) and a numerical simulation of the fundamental mode of the fiber at 1.064\,$\mu$m (right, red arrow denotes the polarization)
    b) Top: Simulated FWM phase-mismatch for the fundamental fiber mode, for increasing the pressure of argon gas filling the the hollow channels of the fiber. Bottom: experimentally measured spectrum after the SCF, when pump and seed are launched into the fiber.
    c) Pump-power scaling of the signal intensity at constant seed power (blue) and its fit (orange).
    d) Signal spectrum measured for increasing argon pressures. Red dashed line is an EME-simulation. The two magenta circles mark the conjugate to the wavelengths marked in Fig.~\ref{fig:TuningBiphotons}.
   }
    \label{fig:TuningParamAmpl}
\end{figure}

Our SCF has an effective core diameter of 830\,nm, defined by a circle inscribed into the central glass region.  For the parametric amplification experiment (Fig.~\ref{fig:TuningParamAmpl}a), the pump is a 1.064\,$\mu$m passively Q-switched laser (1.8\,ns pulses, 42\,kHz repetition rate). Most of the laser power is used to generate a supercontinuum in a solid core photonic crystal fiber~\cite{Wadsworth04} to serve as a seed. The supercontinuum is long-pass filtered (1.15\,$\mu$m cut-on). Pump and seed are prepared to be co-polarized and overlapping in time and space. Both beams are coupled to a single mode ultra-high NA fiber (UHNA), which is butt-coupled to the $\sim 50$\,cm long SCF inside of a mini gas cell (\cite{Pennetta17,Supplemental} for technical details). The out-coupling end of the SCF was mounted in a standard gas cell, where the pressurized gas could be introduced.

 Using an eigenmode expansion (EME), based on a high resolution scanning electron micrograph of the fiber, we predict a phase-matched FWM signal for the fundamental mode near 0.84\,$\mu$m (Fig.~\ref{fig:TuningParamAmpl}b top).
 The fiber presents two distinct polarization states yielding slightly different phase-matching wavelengths. We choose to work with the polarization parallel to one of the nano-membrane (inset in Fig.~\ref{fig:TuningParamAmpl}a, red arrow marks the direction of the polarization). The corresponding phase-matched signal at $\sim 0.84\,\mu$m (Fig.~\ref{fig:TuningParamAmpl}b, bottom) serves as a reference in order to optimize the coupling of the fundamental mode and the polarization states of the pump and seed as well as their time overlap.
We also performed near-field imaging of the different wavelengths exiting the SCF, as well as polarization analysis of the fields to confirm that the process indeed only involves one fiber mode~\cite{Supplemental}. The signal intensity depends on the pump power (Fig.~\ref{fig:TuningParamAmpl}c) as $\sinh^2G$~\cite{Agrawal}, where $G = g\cdot P_\text{Pump}$ is called the parametric gain. From the fit we find $g_\text{avg}=0.54$\,mW$^{-1}$ in terms of the pump average power ($g_\text{PP}=0.0414$\,W$^{-1}$ for peak power~\cite{Supplemental}).

Filling the SCF hollow channels with argon decreases the frequency detuning of the sidebands, as the argon pressure increases (Fig.~\ref{fig:TuningParamAmpl}b, top). 
This is because the evanescent decay length of the mode into the hollow channels scales with the wavelength. Therefore, the increasing gas refractive index (due to increasing gas pressure and therefore density) affects the propagation constant of the idler more strongly than the signal. This tuning mechanism has been used to shift the resonant third harmonic generation (THG) wavelength in sub-micron tapers~\cite{Hammer18}.

By increasing the pressure from 0 to 25\,bar, we observed a continuous shift in the signal wavelength by about 17\,nm ($\sim7$\,THz) towards smaller detuning (Fig.~\ref{fig:TuningParamAmpl}d). The red dashed line corresponds to an EME-simulation. The slight deviation of the simulation from the measured data is mainly due to the neglected chromatic dispersion of the filling gas, an assumption that breaks down at higher gas pressures when chromatic dispersion effects become comparable to the increase in refractive index due to increasing gas density.
Fitting the experimental peak-wavelength of the signal spectrum with a first order polynomial yields an average tuning rate of 0.70\,nm/bar (0.30\,THz/bar). We note that dispersion tuning in SCF has also been demonstrated by depositing a nanofilm on the exposed core~\cite{Warren-Smith17}, however this mechanism is not reversible in contrast to the gas-pressure filling.

\begin{figure}[ht]
    \centering
    \includegraphics[width=\linewidth]{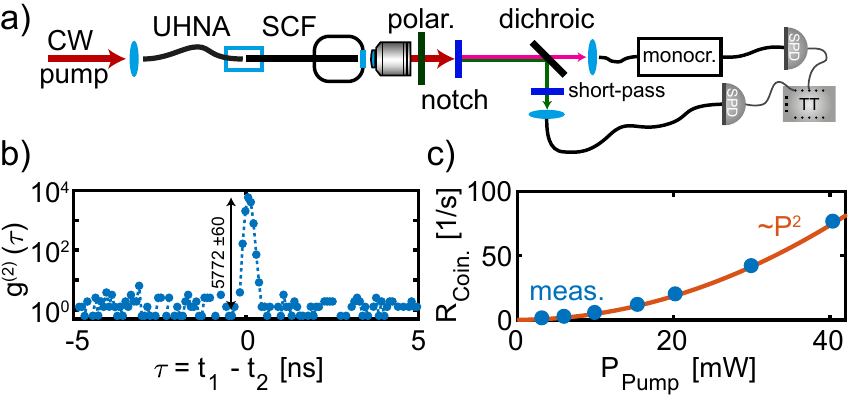}
    \caption{Spontaneous FWM. a) Experimental setup for measuring two-photon coincidences. A CW-pump at $1.064\,\mu$m is launched into the UHNA-SCF device. After re-collimation, light passes through a polarizer and the pump is suppressed using a notch filter. Signal and idler photons are split on a dichroic mirror. Signal photons pass through another set of interference filters, and are coupled to a single-photon detector (SPD). Idler photons pass through a monochromator and are then sent to another SPD. Both SPDs are connected to a time-tagger. b) Measured second-order correlation function. c) Measured coincidence rate for increasing pump powers. Orange line is a quadratic fit. }
    \label{fig:SpontFWM}
\end{figure}
We now consider the spontaneous FWM regime, and demonstrate the effect of the gas pressure on biphoton generation. For these experiments the pump was a 1.064\,$\mu$m continuous wave (CW) laser and no seeding was used. The pump was launched in the same polarization state as found optimal in the previous experiments (Fig.~\ref{fig:SpontFWM}a). At exit from SCF, light passed through a polarizer to select the desired polarization state. The pump was suppressed by a notch filter (-60\,dB), and the sidebands were split on a dichroic mirror. Signal photons passed through another set of interference filters (short pass, 0.9\,$\mu$m cut off) to further suppress the pump, and were then fiber-coupled to a superconducting nanowire single-photon detector (SPD). Idler photons were directed to a fiber-coupled, home-built grating monochromator, and sent to another SPD.
The passband width of the idler monochromator was $\sim$3\,nm. The SPDs were connected to a computer-controlled time-tagger, registering single photon events in either signal or idler channel, as well as coincidences between the two. We used a coincidence resolution time (bin-width) $T_\text{b} = 80$\,ps for the experiments.
Pair generation was characterized by measuring the normalized second-order correlation function (Fig.~\ref{fig:SpontFWM}b), calculated as
\begin{equation}
    g^{(2)}(\tau) = \frac{R_\text{coin}(\tau)}{R_\text{S} R_\text{I} T_\text{b}} ~,
\end{equation}
where $R_\text{coin}(\tau)$ denotes the coincidence rate at a given time delay between two events $\tau=t_1-t_2$ and $R_\text{S,I}$ are the count rates of signal and idler detectors respectively. The pump power was 10\,mW coupled to the fiber, and the monochromator was set to 1.46\,$\mu$m. The peak value of $g^{(2)}$, sometimes called coincidences-to-accidentals ratio (CAR), is $5770\pm60$. Outside of the central peak, $g^{(2)}(\tau)$ drops to  unity -i.e. only accidental coincidences are detected~\footnote{The seemingly high fluctuations in this region are due to the low count rates for these time bins.}. For lower pump powers we measured even higher values of the CAR exceeding $10^4$~\cite{Supplemental}. 
In Fig.~\ref{fig:SpontFWM}c the measured maximum coincidence rate [i.e. the value of $R_\text{coin}(\tau=0)$] is shown as a function of the coupled CW pump power. This includes detection and coupling losses after SCF. The fit in Fig.~\ref{fig:SpontFWM}c is a parabola, corresponding to the low gain expansion of the theoretical power scaling. Through comparison with the previously found value of $g_\text{PP}$ we can estimate the parametric gain as $G\approx4 \cdot 10^{-4}$ at 10\,mW pump power.

The major background to the biphotons stems from spontaneous Raman scattering. While the sideband detuning is larger than the typical extent of the fused silica Raman spectrum, photons from concatenated Raman processes can reach the idler wavelength. Cooling the fiber has been demonstrated as an effective way to reduce these scattering events \cite{Dyer08,Dyer09}.

\begin{figure}[h]
    \centering
    \includegraphics[width=\linewidth]{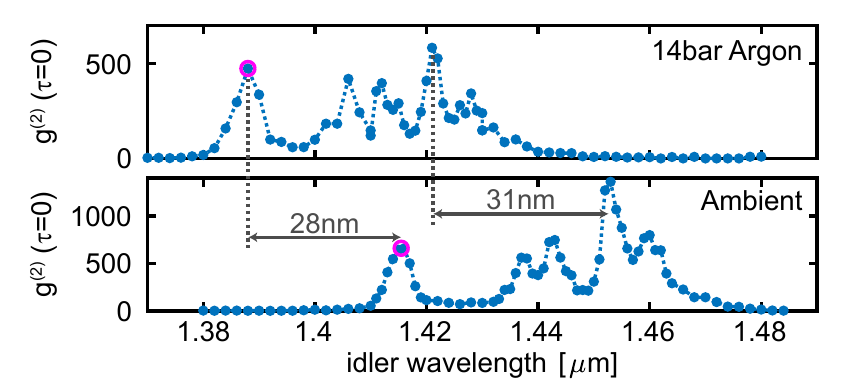}
    \caption{Tuning of spontaneous FWM. Peak value of the second-order correlation function  measured versus the idler wavelength for SCF at ambient pressure (bottom) and filled with 14\,bar argon (top). The spectral features shift by about 30\,nm. The two magenta circles are the idler wavelengths corresponding to the marked wavelengths in Fig.~\ref{fig:TuningParamAmpl}.}
    \label{fig:TuningBiphotons}
\end{figure}
To study the effect of gas-pressure tuning on the spontaneous FWM we performed a set of measurements, where the SCF was either in air (ambient pressure) or filled with 14\,bar of argon. In each measurement the pump power was 20\,mW, and we scanned the monochromator across the spectral region where biphoton emission is expected. At each monochromator setting we collected data for 60\,s.  The CAR versus the monochromator setting is shown in Fig.~\ref{fig:TuningBiphotons}, for the case of air and 14\,bar argon in the SCF. Similar to the seeded case, with increasing argon pressure the emission spectrum shifted to smaller detunings with the same tuning rate ($\sim0.3$\,THz/bar).

The $g^{(2)}$ spectrum demonstrates features very similar to those obtained in the seeded case. In particular, the two magenta markers in Fig.~\ref{fig:TuningBiphotons} correspond to the conjugates of the signal wavelengths that were marked in Fig.~\ref{fig:TuningParamAmpl}c at the same pressures.  The normalized second-order correlation function measured at 14\,bar is slightly reduced compared to the measurement at ambient pressure, possibly due to a higher parametric gain and higher count rates.

\begin{figure}[h]
    \centering
    \includegraphics[width=\linewidth]{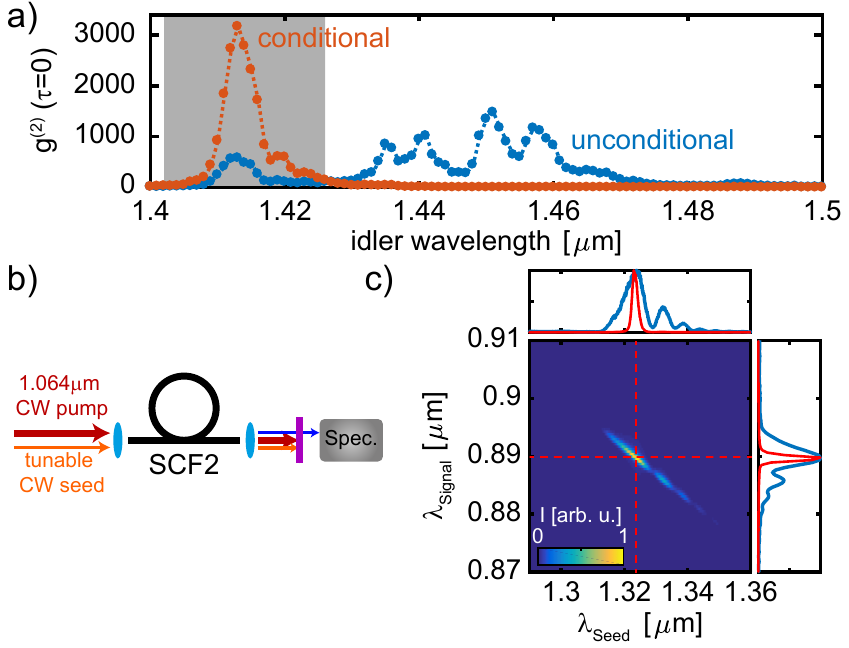}
    \caption{Frequency entanglement. a) Peak value of $g^{(2)}$ as a function of the monochromator wavelength measured at ambient pressure. For the unconditional measurement (blue) the signal photons are unfiltered, whereas for the conditional measurement (orange) a bandpass filter (0.85\,$\mu$m) is inserted in front of the signal SPD. The grey box corresponds to the conjugate of the 0.85\,$\mu$m bandpass full width at half maximum region.
    b) Setup for a stimulated emission tomography (SET) measurement of SCF2 (different phasematching point compared to SCF used before, length $\sim12$\,cm).
    c) SET measurement result (heatmap). The line plots (top and right) are the marginal distributions (blue) and cuts along the dashed lines (red).
    }
    \label{fig:SET}
\end{figure}
We verified the frequency entanglement of the photon pairs comparing conditional and unconditional two-coincidence measurements. In Fig.~\ref{fig:SET}a the peak value of $g^{(2)}$ is plotted as a function of the monochromator wavelength, the experimental conditions were similar to the ambient case in Fig.~\ref{fig:TuningBiphotons}. The blue dots in Fig.~\ref{fig:SET}a correspond to a measurement where all signal photons are directed to the SPD (unconditional measurement). The orange dots correspond to a conditional measurement, i.e. only photons of a specific wavelength (0.85\,$\mu$m) reach the signal SPD. The spectrum of coincidences is much narrower in this scenario, a typical sign for frequency entanglement of the photon pairs~\cite{Mikhailova08,Brida09}.

Additionally we performed a stimulated emission tomography (SET) measurement. SET retrieves the joint spectral intensity (JSI) of a biphoton source by launching the pump alongside a narrowband tunable seed to the system~\cite{Liscidini13,Fang14}. To match the tuning range of our seeding laser (telecom O-band, $\sim 40$\,MHz linewidth), we used a fiber with slightly smaller effective core diameter and therefore smaller frequency detuning, we denote the fiber SCF2. The fiber shows otherwise very similar properties. The SET setup is shown in Fig.~\ref{fig:SET}b, pump and seed (both CW and co-polarized) are launched into SCF2. At exit from the fiber seed and pump are blocked and only the conjugate signal is recorded on a CCD-spectrometer. The JSI is retrieved by scanning the tunable laser across the phase matching point (Fig.~\ref{fig:SET}c). From the JSI we can confirm a high degree of frequency entanglement. The marginal distributions (blue line plots) and the cuts through the JSI (red line plots) correspond to unconditional and conditional measurements at the respective crossection.

We want to note that the overall shape of the emission spectra measured from our SCF are due to inhomogeneities of the fibers, which we confirmed via cutback measurements. We estimate a deviation of $\pm 2.5$\,nm of the effective core diameter over a fiber length of $\sim50$\,cm~\cite{Supplemental}. This could be potentially improved during fiber fabrication, or by careful selection of a shorter fiber piece with less fluctuations. The overall principle of our tuning mechanism etc. is, however, not affected by this. 

In conclusion, we have demonstrated a tunable source of entangled photons, in which the spontaneous emission spectrum can be shifted by 0.70 nm/bar on the signal side by changing the pressure of the filling gas (argon). Our method allows tuning over a wide range of wavelengths, limited only by the maximum achievable refractive index of the filling gas. For the SCF in our experiments we estimate a sideband shift of $\sim40$\,THz ($0.84\,\mu$m to $0.95\,\mu$m) when filling the fiber with up to  45\,bar of xenon.
In the seeded case such a system might be used to implement an all-fiber tunable parametric amplifier. The use of a solid core, small mode area fiber facilitates the use of low-power pump lasers, which may be of interest for small-footprint devices. Furthermore, the same principle could be used for tuning the phase matching of FWM or parametric down-conversion in other types of exposed core waveguides, e.g. on-chip ridge waveguides.

We acknowledge funding by Deutsche Forschungsgemeinschaft (DFG) (CH-1591/3-1, JO-1090/3-1), and the Max Planck School of Photonics.


\bibliography{main}

\begin{thebibliography}{33}%
\makeatletter
\providecommand \@ifxundefined [1]{%
 \@ifx{#1\undefined}
}%
\providecommand \@ifnum [1]{%
 \ifnum #1\expandafter \@firstoftwo
 \else \expandafter \@secondoftwo
 \fi
}%
\providecommand \@ifx [1]{%
 \ifx #1\expandafter \@firstoftwo
 \else \expandafter \@secondoftwo
 \fi
}%
\providecommand \natexlab [1]{#1}%
\providecommand \enquote  [1]{``#1''}%
\providecommand \bibnamefont  [1]{#1}%
\providecommand \bibfnamefont [1]{#1}%
\providecommand \citenamefont [1]{#1}%
\providecommand \href@noop [0]{\@secondoftwo}%
\providecommand \href [0]{\begingroup \@sanitize@url \@href}%
\providecommand \@href[1]{\@@startlink{#1}\@@href}%
\providecommand \@@href[1]{\endgroup#1\@@endlink}%
\providecommand \@sanitize@url [0]{\catcode `\\12\catcode `\$12\catcode
  `\&12\catcode `\#12\catcode `\^12\catcode `\_12\catcode `\%12\relax}%
\providecommand \@@startlink[1]{}%
\providecommand \@@endlink[0]{}%
\providecommand \url  [0]{\begingroup\@sanitize@url \@url }%
\providecommand \@url [1]{\endgroup\@href {#1}{\urlprefix }}%
\providecommand \urlprefix  [0]{URL }%
\providecommand \Eprint [0]{\href }%
\providecommand \doibase [0]{https://doi.org/}%
\providecommand \selectlanguage [0]{\@gobble}%
\providecommand \bibinfo  [0]{\@secondoftwo}%
\providecommand \bibfield  [0]{\@secondoftwo}%
\providecommand \translation [1]{[#1]}%
\providecommand \BibitemOpen [0]{}%
\providecommand \bibitemStop [0]{}%
\providecommand \bibitemNoStop [0]{.\EOS\space}%
\providecommand \EOS [0]{\spacefactor3000\relax}%
\providecommand \BibitemShut  [1]{\csname bibitem#1\endcsname}%
\let\auto@bib@innerbib\@empty
\bibitem [{\citenamefont {Lemieux}\ \emph {et~al.}(2019)\citenamefont
  {Lemieux}, \citenamefont {Giese}, \citenamefont {Fickler}, \citenamefont
  {Chekhova},\ and\ \citenamefont {Boyd}}]{Lemieux2019}%
  \BibitemOpen
  \bibfield  {author} {\bibinfo {author} {\bibfnamefont {S.}~\bibnamefont
  {Lemieux}}, \bibinfo {author} {\bibfnamefont {E.}~\bibnamefont {Giese}},
  \bibinfo {author} {\bibfnamefont {R.}~\bibnamefont {Fickler}}, \bibinfo
  {author} {\bibfnamefont {M.~V.}\ \bibnamefont {Chekhova}},\ and\ \bibinfo
  {author} {\bibfnamefont {R.~W.}\ \bibnamefont {Boyd}},\ }\bibfield  {title}
  {\bibinfo {title} {A primary radiation standard based on quantum nonlinear
  optics},\ }\href {https://doi.org/10.1038/s41567-019-0447-2} {\bibfield
  {journal} {\bibinfo  {journal} {Nature Physics}\ }\textbf {\bibinfo {volume}
  {15}},\ \bibinfo {pages} {529} (\bibinfo {year} {2019})}\BibitemShut
  {NoStop}%
\bibitem [{\citenamefont {F\"ortsch}\ \emph {et~al.}(2013)\citenamefont
  {F\"ortsch}, \citenamefont {F\"urst}, \citenamefont {Wittmann}, \citenamefont
  {Strekalov}, \citenamefont {Aiello}, \citenamefont {Chekhova}, \citenamefont
  {Silberhorn}, \citenamefont {Leuchs},\ and\ \citenamefont
  {Marquardt}}]{Foertsch13}%
  \BibitemOpen
  \bibfield  {author} {\bibinfo {author} {\bibfnamefont {M.}~\bibnamefont
  {F\"ortsch}}, \bibinfo {author} {\bibfnamefont {J.~U.}\ \bibnamefont
  {F\"urst}}, \bibinfo {author} {\bibfnamefont {C.}~\bibnamefont {Wittmann}},
  \bibinfo {author} {\bibfnamefont {D.}~\bibnamefont {Strekalov}}, \bibinfo
  {author} {\bibfnamefont {A.}~\bibnamefont {Aiello}}, \bibinfo {author}
  {\bibfnamefont {M.~V.}\ \bibnamefont {Chekhova}}, \bibinfo {author}
  {\bibfnamefont {C.}~\bibnamefont {Silberhorn}}, \bibinfo {author}
  {\bibfnamefont {G.}~\bibnamefont {Leuchs}},\ and\ \bibinfo {author}
  {\bibfnamefont {C.}~\bibnamefont {Marquardt}},\ }\bibfield  {title} {\bibinfo
  {title} {A versatile source of single photons for quantum information
  processing},\ }\bibfield  {journal} {\bibinfo  {journal} {Nature
  Communications}\ }\textbf {\bibinfo {volume} {4}},\ \href
  {https://doi.org/10.1038/ncomms2838} {10.1038/ncomms2838} (\bibinfo {year}
  {2013})\BibitemShut {NoStop}%
\bibitem [{\citenamefont {Feng}\ \emph {et~al.}(2019)\citenamefont {Feng},
  \citenamefont {Guo},\ and\ \citenamefont {Ren}}]{Feng19}%
  \BibitemOpen
  \bibfield  {author} {\bibinfo {author} {\bibfnamefont {L.-T.}\ \bibnamefont
  {Feng}}, \bibinfo {author} {\bibfnamefont {G.-C.}\ \bibnamefont {Guo}},\ and\
  \bibinfo {author} {\bibfnamefont {X.-F.}\ \bibnamefont {Ren}},\ }\bibfield
  {title} {\bibinfo {title} {Progress on integrated quantum photonic sources
  with silicon},\ }\href {https://doi.org/10.1002/qute.201900058} {\bibfield
  {journal} {\bibinfo  {journal} {Advanced Quantum Technologies}\ ,\ \bibinfo
  {pages} {1900058}} (\bibinfo {year} {2019})}\BibitemShut {NoStop}%
\bibitem [{\citenamefont {Wang}\ \emph {et~al.}(2016)\citenamefont {Wang},
  \citenamefont {Lentine}, \citenamefont {DeRose}, \citenamefont {Starbuck},
  \citenamefont {Trotter}, \citenamefont {Pomerene},\ and\ \citenamefont
  {Mookherjea}}]{Wang16}%
  \BibitemOpen
  \bibfield  {author} {\bibinfo {author} {\bibfnamefont {X.}~\bibnamefont
  {Wang}}, \bibinfo {author} {\bibfnamefont {A.}~\bibnamefont {Lentine}},
  \bibinfo {author} {\bibfnamefont {C.}~\bibnamefont {DeRose}}, \bibinfo
  {author} {\bibfnamefont {A.~L.}\ \bibnamefont {Starbuck}}, \bibinfo {author}
  {\bibfnamefont {D.}~\bibnamefont {Trotter}}, \bibinfo {author} {\bibfnamefont
  {A.}~\bibnamefont {Pomerene}},\ and\ \bibinfo {author} {\bibfnamefont
  {S.}~\bibnamefont {Mookherjea}},\ }\bibfield  {title} {\bibinfo {title}
  {Wide-range and fast thermally-tunable silicon photonic microring resonators
  using the junction field effect},\ }\href
  {https://doi.org/10.1364/oe.24.023081} {\bibfield  {journal} {\bibinfo
  {journal} {Optics Express}\ }\textbf {\bibinfo {volume} {24}},\ \bibinfo
  {pages} {23081} (\bibinfo {year} {2016})}\BibitemShut {NoStop}%
\bibitem [{\citenamefont {Ortiz-Ricardo}\ \emph {et~al.}(2017)\citenamefont
  {Ortiz-Ricardo}, \citenamefont {Bertoni-Ocampo}, \citenamefont
  {Ibarra-Borja}, \citenamefont {Ramirez-Alarcon}, \citenamefont
  {Cruz-Delgado}, \citenamefont {Cruz-Ramirez}, \citenamefont {Garay-Palmett},\
  and\ \citenamefont {U'Ren}}]{Ortiz-Ricardo17}%
  \BibitemOpen
  \bibfield  {author} {\bibinfo {author} {\bibfnamefont {E.}~\bibnamefont
  {Ortiz-Ricardo}}, \bibinfo {author} {\bibfnamefont {C.}~\bibnamefont
  {Bertoni-Ocampo}}, \bibinfo {author} {\bibfnamefont {Z.}~\bibnamefont
  {Ibarra-Borja}}, \bibinfo {author} {\bibfnamefont {R.}~\bibnamefont
  {Ramirez-Alarcon}}, \bibinfo {author} {\bibfnamefont {D.}~\bibnamefont
  {Cruz-Delgado}}, \bibinfo {author} {\bibfnamefont {H.}~\bibnamefont
  {Cruz-Ramirez}}, \bibinfo {author} {\bibfnamefont {K.}~\bibnamefont
  {Garay-Palmett}},\ and\ \bibinfo {author} {\bibfnamefont {A.~B.}\
  \bibnamefont {U'Ren}},\ }\bibfield  {title} {\bibinfo {title} {Spectral
  tunability of two-photon states generated by spontaneous four-wave mixing:
  fibre tapering, temperature variation and longitudinal stress},\ }\href
  {https://doi.org/10.1088/2058-9565/aa7a37} {\bibfield  {journal} {\bibinfo
  {journal} {Quantum Science and Technology}\ }\textbf {\bibinfo {volume}
  {2}},\ \bibinfo {pages} {034015} (\bibinfo {year} {2017})}\BibitemShut
  {NoStop}%
\bibitem [{\citenamefont {Fu}\ \emph {et~al.}(2008)\citenamefont {Fu},
  \citenamefont {Thomas},\ and\ \citenamefont {Dong}}]{Fu08}%
  \BibitemOpen
  \bibfield  {author} {\bibinfo {author} {\bibfnamefont {L.}~\bibnamefont
  {Fu}}, \bibinfo {author} {\bibfnamefont {B.~K.}\ \bibnamefont {Thomas}},\
  and\ \bibinfo {author} {\bibfnamefont {L.}~\bibnamefont {Dong}},\ }\bibfield
  {title} {\bibinfo {title} {Efficient supercontinuum generations in silica
  suspended core fibers},\ }\href {https://doi.org/10.1364/oe.16.019629}
  {\bibfield  {journal} {\bibinfo  {journal} {Optics Express}\ }\textbf
  {\bibinfo {volume} {16}},\ \bibinfo {pages} {19629} (\bibinfo {year}
  {2008})}\BibitemShut {NoStop}%
\bibitem [{\citenamefont {Webb}(2007)}]{Webb07}%
  \BibitemOpen
  \bibfield  {author} {\bibinfo {author} {\bibfnamefont {A.~S.}\ \bibnamefont
  {Webb}},\ }\bibfield  {title} {\bibinfo {title} {Suspended-core holey fiber
  for evanescent-field sensing},\ }\href {https://doi.org/10.1117/1.2430505}
  {\bibfield  {journal} {\bibinfo  {journal} {Optical Engineering}\ }\textbf
  {\bibinfo {volume} {46}},\ \bibinfo {pages} {010503} (\bibinfo {year}
  {2007})}\BibitemShut {NoStop}%
\bibitem [{\citenamefont {Cubillas}\ \emph {et~al.}(2013)\citenamefont
  {Cubillas}, \citenamefont {Unterkofler}, \citenamefont {Euser}, \citenamefont
  {Etzold}, \citenamefont {Jones}, \citenamefont {Sadler}, \citenamefont
  {Wasserscheid},\ and\ \citenamefont {Russell}}]{Cubillas13}%
  \BibitemOpen
  \bibfield  {author} {\bibinfo {author} {\bibfnamefont {A.~M.}\ \bibnamefont
  {Cubillas}}, \bibinfo {author} {\bibfnamefont {S.}~\bibnamefont
  {Unterkofler}}, \bibinfo {author} {\bibfnamefont {T.~G.}\ \bibnamefont
  {Euser}}, \bibinfo {author} {\bibfnamefont {B.~J.~M.}\ \bibnamefont
  {Etzold}}, \bibinfo {author} {\bibfnamefont {A.~C.}\ \bibnamefont {Jones}},
  \bibinfo {author} {\bibfnamefont {P.~J.}\ \bibnamefont {Sadler}}, \bibinfo
  {author} {\bibfnamefont {P.}~\bibnamefont {Wasserscheid}},\ and\ \bibinfo
  {author} {\bibfnamefont {P.~{\relax St.J}.}\ \bibnamefont {Russell}},\
  }\bibfield  {title} {\bibinfo {title} {Photonic crystal fibres for chemical
  sensing and photochemistry},\ }\href {https://doi.org/10.1039/c3cs60128e}
  {\bibfield  {journal} {\bibinfo  {journal} {Chem. Soc. Rev.}\ }\textbf
  {\bibinfo {volume} {42}},\ \bibinfo {pages} {8629} (\bibinfo {year}
  {2013})}\BibitemShut {NoStop}%
\bibitem [{\citenamefont {Pocholle}\ \emph {et~al.}(1985)\citenamefont
  {Pocholle}, \citenamefont {Raffy}, \citenamefont {Papuchon},\ and\
  \citenamefont {Desurvire}}]{Pocholle85}%
  \BibitemOpen
  \bibfield  {author} {\bibinfo {author} {\bibfnamefont {J.~P.}\ \bibnamefont
  {Pocholle}}, \bibinfo {author} {\bibfnamefont {J.}~\bibnamefont {Raffy}},
  \bibinfo {author} {\bibfnamefont {M.}~\bibnamefont {Papuchon}},\ and\
  \bibinfo {author} {\bibfnamefont {E.}~\bibnamefont {Desurvire}},\ }\bibfield
  {title} {\bibinfo {title} {Raman and four photon mixing amplification in
  single mode fibers},\ }\href {https://doi.org/10.1117/12.7973536} {\bibfield
  {journal} {\bibinfo  {journal} {Opt. Eng.}\ }\textbf {\bibinfo {volume}
  {24}},\ \bibinfo {pages} {600} (\bibinfo {year} {1985})}\BibitemShut
  {NoStop}%
\bibitem [{\citenamefont {Hansryd}\ \emph {et~al.}(2002)\citenamefont
  {Hansryd}, \citenamefont {Andrekson}, \citenamefont {Westlund}, \citenamefont
  {Li},\ and\ \citenamefont {Hedekvist}}]{Hansryd02}%
  \BibitemOpen
  \bibfield  {author} {\bibinfo {author} {\bibfnamefont {J.}~\bibnamefont
  {Hansryd}}, \bibinfo {author} {\bibfnamefont {P.}~\bibnamefont {Andrekson}},
  \bibinfo {author} {\bibfnamefont {M.}~\bibnamefont {Westlund}}, \bibinfo
  {author} {\bibfnamefont {J.}~\bibnamefont {Li}},\ and\ \bibinfo {author}
  {\bibfnamefont {P.-O.}\ \bibnamefont {Hedekvist}},\ }\bibfield  {title}
  {\bibinfo {title} {Fiber-based optical parametric amplifiers and their
  applications},\ }\href {https://doi.org/10.1109/jstqe.2002.1016354}
  {\bibfield  {journal} {\bibinfo  {journal} {{IEEE} Journal of Selected Topics
  in Quantum Electronics}\ }\textbf {\bibinfo {volume} {8}},\ \bibinfo {pages}
  {506} (\bibinfo {year} {2002})}\BibitemShut {NoStop}%
\bibitem [{\citenamefont {Li}\ \emph {et~al.}(2004)\citenamefont {Li},
  \citenamefont {Chen}, \citenamefont {Voss}, \citenamefont {Sharping},\ and\
  \citenamefont {Kumar}}]{Li04}%
  \BibitemOpen
  \bibfield  {author} {\bibinfo {author} {\bibfnamefont {X.}~\bibnamefont
  {Li}}, \bibinfo {author} {\bibfnamefont {J.}~\bibnamefont {Chen}}, \bibinfo
  {author} {\bibfnamefont {P.}~\bibnamefont {Voss}}, \bibinfo {author}
  {\bibfnamefont {J.}~\bibnamefont {Sharping}},\ and\ \bibinfo {author}
  {\bibfnamefont {P.}~\bibnamefont {Kumar}},\ }\bibfield  {title} {\bibinfo
  {title} {All-fiber photon-pair source for quantum communications: Improved
  generation of correlated photons},\ }\href
  {https://doi.org/10.1364/opex.12.003737} {\bibfield  {journal} {\bibinfo
  {journal} {Optics Express}\ }\textbf {\bibinfo {volume} {12}},\ \bibinfo
  {pages} {3737} (\bibinfo {year} {2004})}\BibitemShut {NoStop}%
\bibitem [{\citenamefont {Takesue}\ and\ \citenamefont
  {Inoue}(2004)}]{Takesue04}%
  \BibitemOpen
  \bibfield  {author} {\bibinfo {author} {\bibfnamefont {H.}~\bibnamefont
  {Takesue}}\ and\ \bibinfo {author} {\bibfnamefont {K.}~\bibnamefont
  {Inoue}},\ }\bibfield  {title} {\bibinfo {title} {Generation of
  polarization-entangled photon pairs and violation of bell's inequality using
  spontaneous four-wave mixing in a fiber loop},\ }\href
  {https://doi.org/10.1103/physreva.70.031802} {\bibfield  {journal} {\bibinfo
  {journal} {Phys. Rev. A}\ }\textbf {\bibinfo {volume} {70}},\ \bibinfo
  {pages} {031802} (\bibinfo {year} {2004})}\BibitemShut {NoStop}%
\bibitem [{\citenamefont {Rarity}\ \emph {et~al.}(2005)\citenamefont {Rarity},
  \citenamefont {Fulconis}, \citenamefont {Duligall}, \citenamefont
  {Wadsworth},\ and\ \citenamefont {Russell}}]{Rarity05}%
  \BibitemOpen
  \bibfield  {author} {\bibinfo {author} {\bibfnamefont {J.~G.}\ \bibnamefont
  {Rarity}}, \bibinfo {author} {\bibfnamefont {J.}~\bibnamefont {Fulconis}},
  \bibinfo {author} {\bibfnamefont {J.}~\bibnamefont {Duligall}}, \bibinfo
  {author} {\bibfnamefont {W.~J.}\ \bibnamefont {Wadsworth}},\ and\ \bibinfo
  {author} {\bibfnamefont {P.~{\relax St.J}.}\ \bibnamefont {Russell}},\
  }\bibfield  {title} {\bibinfo {title} {Photonic crystal fiber source of
  correlated photon pairs},\ }\href {https://doi.org/10.1364/OPEX.13.000534}
  {\bibfield  {journal} {\bibinfo  {journal} {Optics Express}\ }\textbf
  {\bibinfo {volume} {13}},\ \bibinfo {pages} {534} (\bibinfo {year}
  {2005})}\BibitemShut {NoStop}%
\bibitem [{\citenamefont {Fan}\ \emph {et~al.}(2005)\citenamefont {Fan},
  \citenamefont {Migdall},\ and\ \citenamefont {Wang}}]{Fan05}%
  \BibitemOpen
  \bibfield  {author} {\bibinfo {author} {\bibfnamefont {J.}~\bibnamefont
  {Fan}}, \bibinfo {author} {\bibfnamefont {A.}~\bibnamefont {Migdall}},\ and\
  \bibinfo {author} {\bibfnamefont {L.~J.}\ \bibnamefont {Wang}},\ }\bibfield
  {title} {\bibinfo {title} {Efficient generation of correlated photon pairs in
  a microstructure fiber},\ }\href {https://doi.org/10.1364/ol.30.003368}
  {\bibfield  {journal} {\bibinfo  {journal} {Optics Letters}\ }\textbf
  {\bibinfo {volume} {30}},\ \bibinfo {pages} {3368} (\bibinfo {year}
  {2005})}\BibitemShut {NoStop}%
\bibitem [{\citenamefont {Smith}\ \emph {et~al.}(2009)\citenamefont {Smith},
  \citenamefont {Mahou}, \citenamefont {Cohen}, \citenamefont {Lundeen},\ and\
  \citenamefont {Walmsley}}]{Smith09}%
  \BibitemOpen
  \bibfield  {author} {\bibinfo {author} {\bibfnamefont {B.~J.}\ \bibnamefont
  {Smith}}, \bibinfo {author} {\bibfnamefont {P.}~\bibnamefont {Mahou}},
  \bibinfo {author} {\bibfnamefont {O.}~\bibnamefont {Cohen}}, \bibinfo
  {author} {\bibfnamefont {J.~S.}\ \bibnamefont {Lundeen}},\ and\ \bibinfo
  {author} {\bibfnamefont {I.~A.}\ \bibnamefont {Walmsley}},\ }\bibfield
  {title} {\bibinfo {title} {Photon pair generation in birefringent optical
  fibers},\ }\href {https://doi.org/10.1364/oe.17.023589} {\bibfield  {journal}
  {\bibinfo  {journal} {Optics Express}\ }\textbf {\bibinfo {volume} {17}},\
  \bibinfo {pages} {23589} (\bibinfo {year} {2009})}\BibitemShut {NoStop}%
\bibitem [{\citenamefont {Kim}\ \emph {et~al.}(2019)\citenamefont {Kim},
  \citenamefont {Ihn}, \citenamefont {Kim},\ and\ \citenamefont
  {Shin}}]{Kim19}%
  \BibitemOpen
  \bibfield  {author} {\bibinfo {author} {\bibfnamefont {J.-H.}\ \bibnamefont
  {Kim}}, \bibinfo {author} {\bibfnamefont {Y.~S.}\ \bibnamefont {Ihn}},
  \bibinfo {author} {\bibfnamefont {Y.-H.}\ \bibnamefont {Kim}},\ and\ \bibinfo
  {author} {\bibfnamefont {H.}~\bibnamefont {Shin}},\ }\bibfield  {title}
  {\bibinfo {title} {Photon-pair source working in a silicon-based detector
  wavelength range using tapered micro/nanofibers},\ }\href
  {https://doi.org/10.1364/OL.44.000447} {\bibfield  {journal} {\bibinfo
  {journal} {Opt. Lett.}\ }\textbf {\bibinfo {volume} {44}},\ \bibinfo {pages}
  {447} (\bibinfo {year} {2019})}\BibitemShut {NoStop}%
\bibitem [{\citenamefont {Shukhin}\ \emph {et~al.}(2019)\citenamefont
  {Shukhin}, \citenamefont {Keloth}, \citenamefont {Hakuta},\ and\
  \citenamefont {Kalachev}}]{shukhin19}%
  \BibitemOpen
  \bibfield  {author} {\bibinfo {author} {\bibfnamefont {A.~A.}\ \bibnamefont
  {Shukhin}}, \bibinfo {author} {\bibfnamefont {J.}~\bibnamefont {Keloth}},
  \bibinfo {author} {\bibfnamefont {K.}~\bibnamefont {Hakuta}},\ and\ \bibinfo
  {author} {\bibfnamefont {A.~A.}\ \bibnamefont {Kalachev}},\ }\href@noop {}
  {\bibinfo {title} {Heralded single photon and correlated photon pair
  generation via spontaneous four-wave mixing in tapered optical fibers}}
  (\bibinfo {year} {2019}),\ \Eprint {https://arxiv.org/abs/1910.12918}
  {arXiv:1910.12918 [quant-ph]} \BibitemShut {NoStop}%
\bibitem [{\citenamefont {Finger}\ \emph {et~al.}(2015)\citenamefont {Finger},
  \citenamefont {Iskhakov}, \citenamefont {Joly}, \citenamefont {Chekhova},\
  and\ \citenamefont {Russell}}]{Finger15}%
  \BibitemOpen
  \bibfield  {author} {\bibinfo {author} {\bibfnamefont {M.~A.}\ \bibnamefont
  {Finger}}, \bibinfo {author} {\bibfnamefont {T.~S.}\ \bibnamefont
  {Iskhakov}}, \bibinfo {author} {\bibfnamefont {N.~Y.}\ \bibnamefont {Joly}},
  \bibinfo {author} {\bibfnamefont {M.~V.}\ \bibnamefont {Chekhova}},\ and\
  \bibinfo {author} {\bibfnamefont {P.~{\relax St.J}.}\ \bibnamefont
  {Russell}},\ }\bibfield  {title} {\bibinfo {title} {{Raman-Free,
  Noble-Gas-Filled Photonic-Crystal Fiber Source for Ultrafast, Very Bright
  Twin-Beam Squeezed Vacuum}},\ }\href
  {https://doi.org/10.1103/PhysRevLett.115.143602} {\bibfield  {journal}
  {\bibinfo  {journal} {Phys. Rev. Lett.}\ }\textbf {\bibinfo {volume} {115}},\
  \bibinfo {pages} {143602} (\bibinfo {year} {2015})}\BibitemShut {NoStop}%
\bibitem [{\citenamefont {Finger}\ \emph {et~al.}(2017)\citenamefont {Finger},
  \citenamefont {Joly}, \citenamefont {Russell},\ and\ \citenamefont
  {Chekhova}}]{Finger17}%
  \BibitemOpen
  \bibfield  {author} {\bibinfo {author} {\bibfnamefont {M.~A.}\ \bibnamefont
  {Finger}}, \bibinfo {author} {\bibfnamefont {N.~Y.}\ \bibnamefont {Joly}},
  \bibinfo {author} {\bibfnamefont {P.~S.~J.}\ \bibnamefont {Russell}},\ and\
  \bibinfo {author} {\bibfnamefont {M.~V.}\ \bibnamefont {Chekhova}},\
  }\bibfield  {title} {\bibinfo {title} {Characterization and shaping of the
  time-frequency schmidt mode spectrum of bright twin beams generated in
  gas-filled hollow-core photonic crystal fibers},\ }\href
  {https://doi.org/10.1103/PhysRevA.95.053814} {\bibfield  {journal} {\bibinfo
  {journal} {Phys. Rev. A}\ }\textbf {\bibinfo {volume} {95}},\ \bibinfo
  {pages} {053814} (\bibinfo {year} {2017})}\BibitemShut {NoStop}%
\bibitem [{\citenamefont {Cordier}\ \emph {et~al.}(2019)\citenamefont
  {Cordier}, \citenamefont {Orieux}, \citenamefont {Debord}, \citenamefont
  {G\'{e}rome}, \citenamefont {Gorse}, \citenamefont {Chafer}, \citenamefont
  {Diamanti}, \citenamefont {Delaye}, \citenamefont {Benabid},\ and\
  \citenamefont {Zaquine}}]{Cordier19}%
  \BibitemOpen
  \bibfield  {author} {\bibinfo {author} {\bibfnamefont {M.}~\bibnamefont
  {Cordier}}, \bibinfo {author} {\bibfnamefont {A.}~\bibnamefont {Orieux}},
  \bibinfo {author} {\bibfnamefont {B.}~\bibnamefont {Debord}}, \bibinfo
  {author} {\bibfnamefont {F.}~\bibnamefont {G\'{e}rome}}, \bibinfo {author}
  {\bibfnamefont {A.}~\bibnamefont {Gorse}}, \bibinfo {author} {\bibfnamefont
  {M.}~\bibnamefont {Chafer}}, \bibinfo {author} {\bibfnamefont
  {E.}~\bibnamefont {Diamanti}}, \bibinfo {author} {\bibfnamefont
  {P.}~\bibnamefont {Delaye}}, \bibinfo {author} {\bibfnamefont
  {F.}~\bibnamefont {Benabid}},\ and\ \bibinfo {author} {\bibfnamefont
  {I.}~\bibnamefont {Zaquine}},\ }\bibfield  {title} {\bibinfo {title} {Active
  engineering of four-wave mixing spectral correlations in multiband
  hollow-core fibers},\ }\href {https://doi.org/10.1364/OE.27.009803}
  {\bibfield  {journal} {\bibinfo  {journal} {Opt. Express}\ }\textbf {\bibinfo
  {volume} {27}},\ \bibinfo {pages} {9803} (\bibinfo {year}
  {2019})}\BibitemShut {NoStop}%
\bibitem [{\citenamefont {Wadsworth}\ \emph {et~al.}(2004)\citenamefont
  {Wadsworth}, \citenamefont {Joly}, \citenamefont {Knight}, \citenamefont
  {Birks}, \citenamefont {Biancalana},\ and\ \citenamefont
  {Russell}}]{Wadsworth04}%
  \BibitemOpen
  \bibfield  {author} {\bibinfo {author} {\bibfnamefont {W.~J.}\ \bibnamefont
  {Wadsworth}}, \bibinfo {author} {\bibfnamefont {N.}~\bibnamefont {Joly}},
  \bibinfo {author} {\bibfnamefont {J.~C.}\ \bibnamefont {Knight}}, \bibinfo
  {author} {\bibfnamefont {T.~A.}\ \bibnamefont {Birks}}, \bibinfo {author}
  {\bibfnamefont {F.}~\bibnamefont {Biancalana}},\ and\ \bibinfo {author}
  {\bibfnamefont {P.~{\relax St.J}.}\ \bibnamefont {Russell}},\ }\bibfield
  {title} {\bibinfo {title} {Supercontinuum and four-wave mixing with
  q-switched pulses in endlessly single-mode photonic crystal fibres},\ }\href
  {https://doi.org/10.1364/OPEX.12.000299} {\bibfield  {journal} {\bibinfo
  {journal} {Opt. Express}\ }\textbf {\bibinfo {volume} {12}},\ \bibinfo
  {pages} {299} (\bibinfo {year} {2004})}\BibitemShut {NoStop}%
\bibitem [{\citenamefont {Pennetta}\ \emph {et~al.}(2017)\citenamefont
  {Pennetta}, \citenamefont {Xie}, \citenamefont {Lenahan}, \citenamefont
  {Mridha}, \citenamefont {Novoa},\ and\ \citenamefont {Russell}}]{Pennetta17}%
  \BibitemOpen
  \bibfield  {author} {\bibinfo {author} {\bibfnamefont {R.}~\bibnamefont
  {Pennetta}}, \bibinfo {author} {\bibfnamefont {S.}~\bibnamefont {Xie}},
  \bibinfo {author} {\bibfnamefont {F.}~\bibnamefont {Lenahan}}, \bibinfo
  {author} {\bibfnamefont {M.}~\bibnamefont {Mridha}}, \bibinfo {author}
  {\bibfnamefont {D.}~\bibnamefont {Novoa}},\ and\ \bibinfo {author}
  {\bibfnamefont {P.~{\relax St.J}.}\ \bibnamefont {Russell}},\ }\bibfield
  {title} {\bibinfo {title} {Fresnel-reflection-free self-aligning nanospike
  interface between a step-index fiber and a hollow-core photonic-crystal-fiber
  gas cell},\ }\href {https://doi.org/10.1103/physrevapplied.8.014014}
  {\bibfield  {journal} {\bibinfo  {journal} {Phys. Rev. Appl}\ }\textbf
  {\bibinfo {volume} {8}},\ \bibinfo {pages} {014014} (\bibinfo {year}
  {2017})}\BibitemShut {NoStop}%
\bibitem [{Sup()}]{Supplemental}%
  \BibitemOpen
  \bibfield  {title} {\bibinfo {title} {See supplemental materials: estimation
  of the parametric gain, polarization dependence of the phase matching, origin
  of the shape of the emission spectrum, experimental implementation of the
  mini-cell and measurement of high {CAR}},\ }\href@noop {} {\ }\BibitemShut
  {NoStop}%
\bibitem [{\citenamefont {Agrawal}(2012)}]{Agrawal}%
  \BibitemOpen
  \bibfield  {author} {\bibinfo {author} {\bibfnamefont {G.}~\bibnamefont
  {Agrawal}},\ }\href
  {https://www.elsevier.com/books/nonlinear-fiber-optics/agrawal/978-0-12-397023-7}
  {\emph {\bibinfo {title} {Nonlinear Fiber Optics}}},\ \bibinfo {edition}
  {fifth edition}\ ed.\ (\bibinfo  {publisher} {Academic Press},\ \bibinfo
  {year} {2012})\BibitemShut {NoStop}%
\bibitem [{\citenamefont {Hammer}\ \emph {et~al.}(2018)\citenamefont {Hammer},
  \citenamefont {Cavanna}, \citenamefont {Pennetta}, \citenamefont {Chekhova},
  \citenamefont {Russell},\ and\ \citenamefont {Joly}}]{Hammer18}%
  \BibitemOpen
  \bibfield  {author} {\bibinfo {author} {\bibfnamefont {J.}~\bibnamefont
  {Hammer}}, \bibinfo {author} {\bibfnamefont {A.}~\bibnamefont {Cavanna}},
  \bibinfo {author} {\bibfnamefont {R.}~\bibnamefont {Pennetta}}, \bibinfo
  {author} {\bibfnamefont {M.~V.}\ \bibnamefont {Chekhova}}, \bibinfo {author}
  {\bibfnamefont {P.~{\relax St.J}.}\ \bibnamefont {Russell}},\ and\ \bibinfo
  {author} {\bibfnamefont {N.~Y.}\ \bibnamefont {Joly}},\ }\bibfield  {title}
  {\bibinfo {title} {Dispersion tuning in sub-micron tapers for third-harmonic
  and photon triplet generation},\ }\href
  {https://doi.org/10.1364/OL.43.002320} {\bibfield  {journal} {\bibinfo
  {journal} {Optics Letters}\ }\textbf {\bibinfo {volume} {43}},\ \bibinfo
  {pages} {2320} (\bibinfo {year} {2018})}\BibitemShut {NoStop}%
\bibitem [{\citenamefont {Warren-Smith}\ \emph {et~al.}(2017)\citenamefont
  {Warren-Smith}, \citenamefont {Chemnitz}, \citenamefont {Schneidewind},
  \citenamefont {Kostecki}, \citenamefont {Ebendorff-Heidepriem}, \citenamefont
  {Monro},\ and\ \citenamefont {Schmidt}}]{Warren-Smith17}%
  \BibitemOpen
  \bibfield  {author} {\bibinfo {author} {\bibfnamefont {S.~C.}\ \bibnamefont
  {Warren-Smith}}, \bibinfo {author} {\bibfnamefont {M.}~\bibnamefont
  {Chemnitz}}, \bibinfo {author} {\bibfnamefont {H.}~\bibnamefont
  {Schneidewind}}, \bibinfo {author} {\bibfnamefont {R.}~\bibnamefont
  {Kostecki}}, \bibinfo {author} {\bibfnamefont {H.}~\bibnamefont
  {Ebendorff-Heidepriem}}, \bibinfo {author} {\bibfnamefont {T.~M.}\
  \bibnamefont {Monro}},\ and\ \bibinfo {author} {\bibfnamefont {M.~A.}\
  \bibnamefont {Schmidt}},\ }\bibfield  {title} {\bibinfo {title}
  {Nanofilm-induced spectral tuning of third harmonic generation},\ }\href
  {https://doi.org/10.1364/OL.42.001812} {\bibfield  {journal} {\bibinfo
  {journal} {Optics Letters}\ }\textbf {\bibinfo {volume} {42}},\ \bibinfo
  {pages} {1812} (\bibinfo {year} {2017})}\BibitemShut {NoStop}%
\bibitem [{Note1()}]{Note1}%
  \BibitemOpen
  \bibinfo {note} {The seemingly high fluctuations in this region are due to
  the low count rates for these time bins.}\BibitemShut {Stop}%
\bibitem [{\citenamefont {Dyer}\ \emph {et~al.}(2008)\citenamefont {Dyer},
  \citenamefont {Stevens}, \citenamefont {Baek},\ and\ \citenamefont
  {Nam}}]{Dyer08}%
  \BibitemOpen
  \bibfield  {author} {\bibinfo {author} {\bibfnamefont {S.~D.}\ \bibnamefont
  {Dyer}}, \bibinfo {author} {\bibfnamefont {M.~J.}\ \bibnamefont {Stevens}},
  \bibinfo {author} {\bibfnamefont {B.}~\bibnamefont {Baek}},\ and\ \bibinfo
  {author} {\bibfnamefont {S.~W.}\ \bibnamefont {Nam}},\ }\bibfield  {title}
  {\bibinfo {title} {High-efficiency, ultra low-noise all-fiber photon-pair
  source},\ }\href {https://doi.org/10.1364/oe.16.009966} {\bibfield  {journal}
  {\bibinfo  {journal} {Optics Express}\ }\textbf {\bibinfo {volume} {16}},\
  \bibinfo {pages} {9966} (\bibinfo {year} {2008})}\BibitemShut {NoStop}%
\bibitem [{\citenamefont {Dyer}\ \emph {et~al.}(2009)\citenamefont {Dyer},
  \citenamefont {Baek},\ and\ \citenamefont {Nam}}]{Dyer09}%
  \BibitemOpen
  \bibfield  {author} {\bibinfo {author} {\bibfnamefont {S.~D.}\ \bibnamefont
  {Dyer}}, \bibinfo {author} {\bibfnamefont {B.}~\bibnamefont {Baek}},\ and\
  \bibinfo {author} {\bibfnamefont {S.~W.}\ \bibnamefont {Nam}},\ }\bibfield
  {title} {\bibinfo {title} {High-brightness, low-noise, all-fiber photon pair
  source},\ }\href {https://doi.org/10.1364/OE.17.010290} {\bibfield  {journal}
  {\bibinfo  {journal} {Optics Express}\ }\textbf {\bibinfo {volume} {17}},\
  \bibinfo {pages} {10290} (\bibinfo {year} {2009})}\BibitemShut {NoStop}%
\bibitem [{\citenamefont {Mikhailova}\ \emph {et~al.}(2008)\citenamefont
  {Mikhailova}, \citenamefont {Volkov},\ and\ \citenamefont
  {Fedorov}}]{Mikhailova08}%
  \BibitemOpen
  \bibfield  {author} {\bibinfo {author} {\bibfnamefont {Y.~M.}\ \bibnamefont
  {Mikhailova}}, \bibinfo {author} {\bibfnamefont {P.~A.}\ \bibnamefont
  {Volkov}},\ and\ \bibinfo {author} {\bibfnamefont {M.~V.}\ \bibnamefont
  {Fedorov}},\ }\bibfield  {title} {\bibinfo {title} {Biphoton wave packets in
  parametric down-conversion: Spectral and temporal structure and degree of
  entanglement},\ }\href {https://doi.org/10.1103/physreva.78.062327}
  {\bibfield  {journal} {\bibinfo  {journal} {Physical Review A}\ }\textbf
  {\bibinfo {volume} {78}},\ \bibinfo {pages} {062327} (\bibinfo {year}
  {2008})}\BibitemShut {NoStop}%
\bibitem [{\citenamefont {Brida}\ \emph {et~al.}(2009)\citenamefont {Brida},
  \citenamefont {Caricato}, \citenamefont {Fedorov}, \citenamefont {Genovese},
  \citenamefont {Gramegna},\ and\ \citenamefont {Kulik}}]{Brida09}%
  \BibitemOpen
  \bibfield  {author} {\bibinfo {author} {\bibfnamefont {G.}~\bibnamefont
  {Brida}}, \bibinfo {author} {\bibfnamefont {V.}~\bibnamefont {Caricato}},
  \bibinfo {author} {\bibfnamefont {M.~V.}\ \bibnamefont {Fedorov}}, \bibinfo
  {author} {\bibfnamefont {M.}~\bibnamefont {Genovese}}, \bibinfo {author}
  {\bibfnamefont {M.}~\bibnamefont {Gramegna}},\ and\ \bibinfo {author}
  {\bibfnamefont {S.~P.}\ \bibnamefont {Kulik}},\ }\bibfield  {title} {\bibinfo
  {title} {Characterization of spectral entanglement of spontaneous
  parametric-down conversion biphotons in femtosecond pulsed regime},\ }\href
  {https://doi.org/10.1209/0295-5075/87/64003} {\bibfield  {journal} {\bibinfo
  {journal} {{EPL} (Europhysics Letters)}\ }\textbf {\bibinfo {volume} {87}},\
  \bibinfo {pages} {64003} (\bibinfo {year} {2009})}\BibitemShut {NoStop}%
\bibitem [{\citenamefont {Liscidini}\ and\ \citenamefont
  {Sipe}(2013)}]{Liscidini13}%
  \BibitemOpen
  \bibfield  {author} {\bibinfo {author} {\bibfnamefont {M.}~\bibnamefont
  {Liscidini}}\ and\ \bibinfo {author} {\bibfnamefont {J.~E.}\ \bibnamefont
  {Sipe}},\ }\bibfield  {title} {\bibinfo {title} {{Stimulated Emission
  Tomography}},\ }\href {https://doi.org/10.1103/PhysRevLett.111.193602}
  {\bibfield  {journal} {\bibinfo  {journal} {Physical Review Letters}\
  }\textbf {\bibinfo {volume} {111}},\ \bibinfo {pages} {193602} (\bibinfo
  {year} {2013})}\BibitemShut {NoStop}%
\bibitem [{\citenamefont {Fang}\ \emph {et~al.}(2014)\citenamefont {Fang},
  \citenamefont {Cohen}, \citenamefont {Liscidini}, \citenamefont {Sipe},\ and\
  \citenamefont {Lorenz}}]{Fang14}%
  \BibitemOpen
  \bibfield  {author} {\bibinfo {author} {\bibfnamefont {B.}~\bibnamefont
  {Fang}}, \bibinfo {author} {\bibfnamefont {O.}~\bibnamefont {Cohen}},
  \bibinfo {author} {\bibfnamefont {M.}~\bibnamefont {Liscidini}}, \bibinfo
  {author} {\bibfnamefont {J.~E.}\ \bibnamefont {Sipe}},\ and\ \bibinfo
  {author} {\bibfnamefont {V.~O.}\ \bibnamefont {Lorenz}},\ }\bibfield  {title}
  {\bibinfo {title} {Fast and highly resolved capture of the joint spectral
  density of photon pairs},\ }\href {https://doi.org/10.1364/OPTICA.1.000281}
  {\bibfield  {journal} {\bibinfo  {journal} {Optica}\ }\textbf {\bibinfo
  {volume} {1}},\ \bibinfo {pages} {281} (\bibinfo {year} {2014})}\BibitemShut
  {NoStop}%
\end{thebibliography}%

\end{document}